\documentclass[oldversion]{aa}

\usepackage{epsfig}
\usepackage{graphics}
\usepackage{float}
\usepackage{amsmath}
\usepackage{multirow}
\usepackage{longtable}
\usepackage{rotate}
\usepackage{array}
\usepackage{subfigure}
\def\kms{\ifmmode{\rm km\,s^{-1}}\else\hbox{$\rm km\,s^{-1}$}\fi}

\setlongtables

\begin{document}

\title{Irradiated stars with convective envelopes}

\author{L.B.Lucy}
\offprints{L.B.Lucy}
\institute{Astrophysics Group, Blackett Laboratory, Imperial College 
London, Prince Consort Road, London SW7 2AZ, UK}
\date{Received ; Accepted }

\abstract{The observed radii of M dwarfs in eclipsing binaries exceed 
predicted radii by $\sim 5\%$. To investigate this anomaly, 
the structure of low-mass stars irradiated by a close companion
is considered. Irradiation modifies
the surface boundary conditions and thereby also the adiabatic constants 
of their outer convection zones.
This changes the models' radii and luminosities.
For short-period M dwarf binaries with components of similar mass, 
radius inflation due to their mutual irradiation is found to be
$ \la 0.4\%$. This is an order of magnitude too small to explain the
observed anomaly. Although
stronger irradiation results in a monotonically increasing 
radius, a saturation effect limits the inflation to $\la 5\%$.   
}
\authorrunning{Lucy}
\titlerunning{Irradiated stars}
\maketitle
%
%
%
\section{Introduction}
A persistent problem for our understanding of M dwarfs is that 
radii derived from the components of eclipsing binaries are larger 
than radii predicted by stellar models. In a review from an observer's
perspective, Torres (2013) reports discrepancies 
of $5-10\%$ for the best measured systems.  
Somewhat smaller discrepancies $ \sim 3\%$ are reported by
Spada et al. (2013) when the data are compared to their newly
computed evolutionary tracks for low-mass stars. Nevertheless, Spada et al.
emphasize that the existence of the discrepacy is beyond doubt.

An important clue to the cause of inflated radii is that
the discrepacy is larger for the components of short period
binaries (Kraus et al 2010; Spada et al. 2013). This has prompted the
seemingly plausible suggestion 
(e.g.,  Lopez-Morales 2007; Chabrier, Gallardo \& Baraffe 2007;  MacDonald \& Mullan 2013)
that enhanced dynamo activity due to
synchronized rotation leads to stronger magnetic fields and hence inhibited   
convective energy transport in these star's envelopes. 
But since
a treatment of this effect from first principles is not yet feasible,
a defintive test is not possible. Moreover, recent papers 
(Feiden \& Chaboyer
2014; Browning et al. 2016) question the feasibility of the very
strong fields required to achieve the observed inflation. 

In this paper, another seemingly plausible 
possibility is
investigated, namely that inflated radii in short-period low-mass binaries
are due to the components' mutual irradiation. An attractive aspect of this
suggestion is that a treatment from first principles is not beyond our
current theoretical understanding.
\section{The Model}
A general treatment of the interaction  
of a close binary's components 
requires the simultaneous solution of their rotationally- and tidally-
distorted structures, with irradiation by each component incorporated into 
the photospheric boundary conditions of its companion.
Although the relevant aspects of the theories of stellar structure 
and stellar atmospheres are well understood, this general 
problem is a formidable technical challenge.
Accordingly, a severely simplified, tractable version is now defined.

The adopted model is illustrated in Fig.1: A non-rotating, zero age
main sequence (ZAMS) star
is irradiated by a point source of luminosity $L_{\star}$ at  
separation $a$. In addition, the point source is massless and is {\em not} in
orbit around the target star.  

In adopting this model, rotational- and tidal distortions are eliminated, 
as is the complicating effect of mutual irradiation.
Because of these simplifications, this model isolates the
{\em differential} effect of irradiation.

Historically, discussions of irradiation in close binaries
concern light curve perturbations (``the reflection effect'') 
and their impact on  
photometric elements (Eddington 1926; Milne 1926). Because of the
fundamental importance of binary star data, this aspect of irradiation
is now highly developed (e.g., Wilson 1990). Here a simpler treatment is
adopted in order to incorporate irradiation into stellar structure
calculations. 

%
%
\begin{figure}
\vspace{8.2cm}
\includegraphics{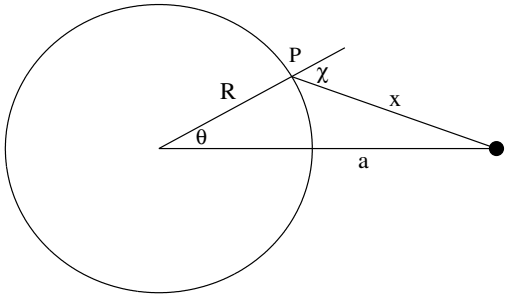}
\caption{Irradiation model. A non-rotating star of radius $R$ is irradiated 
by a stationary point source of luminosity $L_{\star}$ at separation $a$. 
The critical angle 
$\theta_{t}$ separating the irradiated and dark segments is given by 
$\cos \theta_{t} = \upsilon_{t} = R/a$, corresponding to irradiation at grazing 
incidence ($\chi = \pi/2$).} 
\end{figure}
\subsection{Model segments}
Because the irradiating source is a point, there is a sharply defined 
terminator separating the irradiated and dark sides of the target star 
- i.e., there is no penumbral zone.
In terms of polar coordinates with the point source at $z = a$, the 
terminator is the circle $r = R, \: \theta = \theta_{t}$, where
$R$ is the star's radius, and $\cos \theta_{t} = \upsilon_{t} =  R/a$.

In addition to dividing the surface into two zones, we can 
conveniently use it to divide the entire structure into two 
segments: 1)
an {\em irradiated} segment comprising all elements of the star with
$r \le R$ and $\theta \le \theta_{t}$; and 2) a {\em dark} segment 
comprising all elements with $r \le R$ and $\theta > \theta_{t}$.

Thus defined, the irradiated and dark segments comprise the fractions
$1-\zeta$ and $\zeta$ of the star's total surface area and volume,
where
\begin{equation}
  \zeta = \frac{1 + \upsilon_{t}}{2}  \;\;\; with \;\;\;
                                  \upsilon_{t} = \frac{R}{a}
\end{equation}
Moreover, the conical contact surface between the two segments is defined by
$ r \in (0,R), \: \theta = \theta_{t}$.
%
%
%
%
\section{Basic code}
The starting point for the calculations reported in this paper is
a standard Henyey stellar structure code for ZAMS stars. 
\subsection{Input physics}
A brief summary of the input physics is as follows: 

The composition is $X=0.70, Z=0.02$. A table of opacities (Rosseland mean
coefficients)
for this standard Population I composition was constucted using data 
downloaded from the 
Livermore (OPAL, Iglesias \& Rogers 1996) and Wichita 
(Alexander \& Ferguson 1994) websites. The Wichita opacities are adopted
for $\log T \le 3.70$ and the OPAL opacities for $\log T > 3.70$.  

Equation of state data was also downloaded from the Livermore
(OPAL, Rogers et al. 1996) website. The data used is interpolated to the 
adopted composition from the OPAL files as updated on 22 Feb 2006.
    
The outer convection zone is treated using the formulation of mixing-length
theory presented by Kippenhahn et al. (2012). To facilitate comparison
with the Yale models (Spada et al. 2013), their solar-calibrated value
of the mixing-length ratio $\alpha = 1.875$ is adopted.

Nuclear energy production also follows the treatment given in 
Kippenhahn et al. (2012).

One effect neglected here but taken into account in the Yale models is the
gravitational settling of helium and heavy elements. Thus ZAMS models
computed with the code described here are strictly homogeneous. 
But this neglect is of little concern since
the code is used to quantify the {\em differential} effect of irradiation
rather than to predict absolute values of fundamental parameters. 
\subsection{Photosphere}
Non-grey effects are important in the atmospheres of M stars and must
be taken into account in the surface boundary conditions 
(e.g. Baraffe et al. 2015, and references therein). 
On the other hand, grey atmosphere theory 
is a valuable simplification when incorporating irradiation.
These conflicting considerations are reconciled by 
adopting grey atmosphere theory but calibrating the absorption
coefficient so as to accurately match the photospheric pressure of
non-grey models.

Let $k_{\rm eff}(T_{\rm eff}, g)$ denote the {\em constant} effective grey 
absorption coefficient within
an atmosphere of effective temperature $T_{\rm eff}$ and surface 
gravity $g$. In radiative equilibrium, the temperature stratification 
as a function of optical depth $\tau$ is well approximated with the
Milne-Eddington formula
\begin{equation}
  T^{4} =  \frac{1}{2} \: T_{\rm eff}^{4} \: (1 + \frac{3}{2} \tau ) 
\end{equation}
In mechanical (hydrostatic) equilibrium, the corresponding pressure 
distribution is given by
\begin{equation}
  P = \frac{g}{k_{\rm eff}} \: \tau 
\end{equation}

A 2-D tabulation of the function $k_{\rm eff}(T_{\rm eff}, g)$ is derived by
fitting to the non-grey stellar atmospheres of Gustafsson et al. (2008). 
Specifically, models with solar
composition and zero microturbulence were downloaded from the MARCS website.
For each model, the layer with temperature $T_{*}$ closest to $T_{\rm eff}$ is
selected, and its Milne-Eddington optical depth $\tau_{*}$
computed from Eq.(2). Then, if $P_{*}$ is the pressure of the selected layer,
Eq.(3) is satisfied if $k_{\rm eff} = g \tau_{*} / P_{*}$. 
With this prescription, if we compute the structure of an atmosphere  
assuming mechanical equilibrium and a Milne-Eddington temperature 
stratification,
the photospheric pressure will be almost exactly that of the corresponding
MARCS model. 

The accuracy of the derived values $k_{\rm eff}$ and the resulting
photospheric pressures can be gauged from corresponding calculations with
the less extensive grid of Castelli \& Kurucz (2004). For 
$T_{\rm eff} \in (3500,5500K)$ and $\log g \in (4.0,5.0)$, the mean absolute
difference in $\log k_{\rm eff} = 0.031$. However, for models at 
$T_{\rm eff} = 3500K$ and $\log g \in (4.0,5.0)$, the difference rises to 
$0.143$.  

Eqns (2) and (3) evaluated at $\tau = 2/3$ dermine the starting point
$(P,T)$ for inward envelope integrations. Effectively,
with $k_{\rm eff}$ calibrated as described above, the starting point 
is derived from the MARCS models. In these inward integrations,
the opacities are from the Wichita-OPAL data and convection is treated with
mixing length theory (Sect.3.1).

\subsection{Model comparisons}
The Yale evolutionary tracks  
discussed in Sect.1 include nuclear burning during contraction to the main 
sequence, and so no point
along these tracks corresponds exactly to classically-defined ZAMS models. 
Approximate ZAMS parameters can, however, be derived by selecting models 
at $5 Gyr$ - see Fig.6 in Spada et al. 2013. These are then compared with ZAMS 
models computed with the code described above. The models agree closely at
${\cal M} \sim 0.40-0.45 {\cal M}_{\sun}$ but less so towards smaller masses.
Thus at ${\cal M} = 0.2 {\cal M}_{\sun}$, the Yale track's radius and 
luminosity are $6$ and $13\%$ smaller, respectively, than the present code's 
ZAMS values.  
\subsection{Sub-photospheric convection}
For later reference, some typical aspects of sub-photosperic convection in
low-mass ZAMS stars are here noted. The ${\cal M} = 0.4 {\cal M}_{\sun}$ model
has $T_{\rm eff} = 3582K$, and its atmosphere becomes convectively unstable
at $T = 3711K$ - i.e., at $\tau = 0.87$. Initially, the convective energy
transport is weak, but with increasing depth quickly becomes efficient.
Thus, $99\%$ of the  energy flux is carried by convection at $T = 7428K$
- i.e., $\tau = 11.7$.

At yet greater depths, convection becomes highly 
efficient; the temperature gradient $\nabla = d \log T/ d \log P$ then only 
slightly exceeds the
adiabatic gradient $\nabla_{ad}$, and so the stratification is 
almost exactly isentropic. The departure from isentropy can be
quantified as 
\begin{equation}
  \eta = \frac{\nabla - \nabla_{ad}}{\nabla_{ad}} 
\end{equation}
Thus at $T = 10^{5}K$, $\eta = 4.4 \times 10^{-6}$. Moreover, convection
becomes 
highly efficient with only a fractionally small penetration
below the surface. Thus, at $r/R = 0.99$, the ratio 
$\eta = 1.1 \times 10^{-5}$. This implies that the entropy constant 
of the convection zone is to an excellent approximation a function of
just $T_{\rm eff}$ and $g$, and so can be derived with a
plane-parallel calculation.   

When convective transfer is efficient,
the nearly exact isentropic stratification is effectively uncoupled from
the total energy flux. Thus, for example, if the luminosity 
variable $L_{r}$ were to be doubled due an injection of
energy at $r$, the expected change in
$\nabla$ is $\sim \eta \nabla_{ad}$, and is thus negligible when 
$\eta \la 10^{-5}$. The convective cores of massive stars exemplify this
point: $L_{r}$ increases with $r$ due to nuclear burning and yet the 
isentropic approximation is superb since $\eta \sim 10^{-6}$ 
(Schwarzschild 1958, p.50).
\section{Structure of the dark segment}
The code described in Section 3 is now used to compute
the structure of the dark segment. 
\subsection{Convection zone}
The coupling of the dark and the irradiated segments of the target star
occurs within their shared sub-photosperic convection zone in a manner
similar to the coupling of the components of late-type contact binaries
(Lucy 1968a). 
Thus, in an investigation (Lucy 1968b) of the light curves of these binaries,
it was remarked that a treatment of the reflection effect requires that inward
integrations of the equations governing atmospheric structure must
all finish on the same adiabat. In the context of non-contact binaries
- especially the secondaries of Algol systems - Rucinski (1969) 
investigated this effect in detail obtaining the important result that
the bolometric albedo is $\sim 0.4-0.5$, thus providing a physical
interpretation of the earlier empirical discovery by Hosokawa (1960).    
      
The implication of a bolometric albedo less than the value unity expected for
a radiative envelope (Eddington 1926) is that a lateral transfer of energy
from the irradiated to the dark segment occurs within the shared convection 
zone. Indeed, Hosokawa inferred ``that a circulation of matter on a large 
scale will occur through the superficial layers.''

Let the rate of energy transfer from the irradiated to the dark segment be
${\cal L}$. This transfer occurs across the contact surface between
the segments. In consequence, the condition of thermal equilibrium for
the dark segment must include this term. If $L_{nuc}$ is the nuclear 
energy generation rate, and if
\begin{equation}
  L_{\dag} =  4\pi R^{2} \times \pi F^{drk}
\end{equation}
where $\pi F^{drk}$ is the emitted flux on the dark segment's surface, 
then the condition
of thermal equilibrium for the dark segment is 
\begin{equation}
  \zeta \: \L_{\dag} = \zeta \: L_{nuc} + {\cal L} 
\end{equation}

Now consider a radius vector within the dark segment - i.e. 
$\theta > \theta_{t}$. Deep in the convection zone below the zone
of lateral energy exchange but exterior to the nuclear burning core,
the luminosity variable $L_{r} = L_{nuc}$. But, as the radius vector
reaches the surface, $L_{r} = L_{\dag}$ - i.e., an increase by
\begin{equation}
  \Delta L = {\cal L} \: / \: \zeta
\end{equation}
Accordingly, in computing the structure of the dark segment, it is 
convenient to define an equivalent single star with luminosity
$L_{\dag}$. But this star is not in thermal equilibrium because 
within its convection zone, $L_{r}$ increases from $L_{nuc}$ 
to $L_{\dag} = L_{nuc} + \Delta L$. 

In the dark segment, the functional form of the increase of $L_{r}$
from $L_{nuc}$ to $L_{\dag}$ will depend on $\theta$ (Fig. 1). 
However, if we suppose that this increase occurs entirely within
the zone where convection is efficient and if we take the isentropic
limit, the functional form of this increase becomes
irrelevant.
\subsection{Luminosities and radii}
To compute the structure of the equivalent single star, we simply
modify the code described in Sect.3 by 
introducing a jump $\Delta L$  in $L_{r}$ 
at any point in the envelope where
convection is efficient. In this way, for a given ${\cal M}$, 
a 1-D sequence of models is created with luminosities 
$L(\Delta L;{\cal M}) = L_{\dag}$ and radii $R(\Delta L;{\cal M})$.
These functions can be computed without specifying $a$ or $L_{\star}$.     

Because of the energy injection, the photospheric
luminosity $L_{\dag}$ of the equivalent 
single star exceeds its nuclear luminosity $L_{nuc}$ by the amount   
$\Delta L = {\cal L}/\zeta$. The divergence of $L_{\dag}$ and $L_{nuc}$
with increasing $\Delta L$ is shown in Fig.2 for 
${\cal M} = 0.4 {\cal M}_{\sun}$. 
A notable feature is that $L_{nuc}$ drops markedly with increasing
$\Delta L$. Thus $L_{nuc}$ drops by a factor of $\sim 2$ when 
$\Delta L$ increases to $L_{0}$, the zero-age luminosity.   
Because of their deep convective envelopes
with isentropic stratification, the internal structures of low-mass stars 
are {\em not} insensitive to the surface boundary condition.
Here irradiation via the input ${\cal L}$ implies a change in the surface 
boundary conditions resulting in a changed adiabat.
%
%
%
%
\begin{figure}
\vspace{8.2cm}
\includegraphics{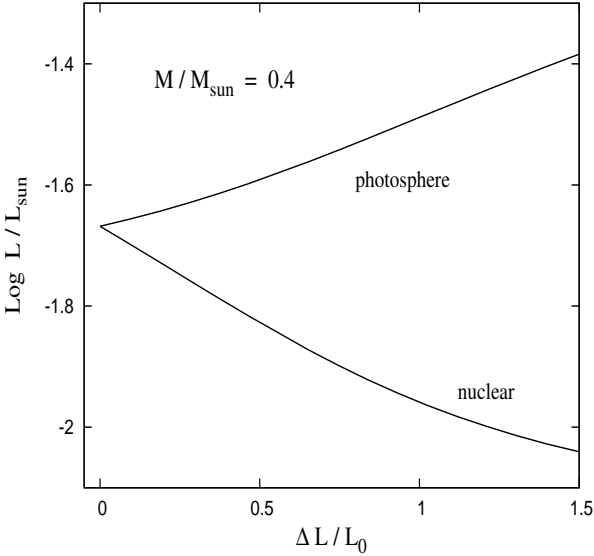}
\caption{Luminosities of the equivalent single star of mass 
$0.4 {\cal M}_{\sun}$ as functions of the parameter $\Delta L / L_{0}$,
where $L_{0}$ is the ZAMS luminosity.
Plotted are the photospheric ($L_{\dag}$) and nuclear ($L_{nuc}$)
luminosities. Note that $L_{\dag} = L_{nuc} + \Delta L$.} 
\end{figure}
The corresponding changes in radii are plotted in Fig.3 for 
${\cal M}/{\cal M}_{\sun} = 0.2(0.1)0.6$. As expected, the effect of 
energy input is to inflate the dark segment's radius.

Because of their
intrinsic interest, the calculations 
reported in Fig.3 explore the inflation effect far beyond inputs
$\Delta L$ relevant for main sequence binaries. Examples of such 
enhanced
inflation may eventually be found for the secondaries in sdB + dM binaries - 
see Heber (2016, Sect. 5.2) for a recent review.   

For M dwarfs in main sequence binaries, radii inflated by up to $10\%$ are
reported (Sect.1). From Fig.3, we find that
the radius of a $0.4 {\cal M}_{\sun}$ ZAMS star is inflated by $5\%$
when $\Delta L = 7.27 \times 10^{-3} L_{\sun}$. But this is 
$ = 0.339 \: L_{0} $,
suggesting
already that $5\%$ inflation is highly unlikely when the companion is a main
sequence
star with mass comparable to that of the target star.        
%
%
%
%
\begin{figure}
\vspace{8.2cm}
\includegraphics{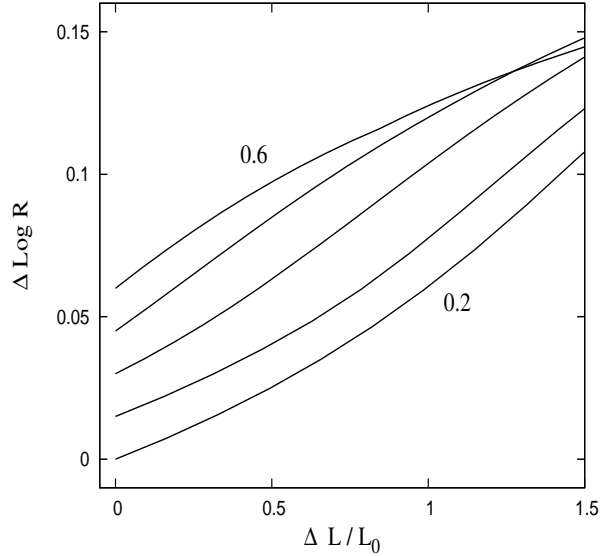}
\caption{Radius inflation due to irradiation. $\Delta log R$ is plotted
against $\Delta L / L_{0}$ for masses ${\cal M} = 0.2(0.1)0.6 {\cal M}_{\sun}$. 
Here $\Delta L$ is the increment in $L_{r}$ within the convective
envelope and $L_{0}$ is the ZAMS luminosity for mass ${\cal M}$. For
clarity, consecutive curves are displaced vertically by $0.015$. } 
\end{figure}
%
%
%
%
%
%
\section{Irradiated segment}
The structure of the irradiated segment is now considered,
with the aim of
relating the energy input parameter $\Delta L = {\cal L}/\zeta$
to the parameters $L_{\star}$ and $a$ that define the irradiating source.

As noted in Sect.4.1, this segment is coupled to the dark segment
via their shared convection zone, and in which the gradient of entropy
is negligible once the convection becomes efficient.
Because this occurs at a fractionally small depth below the surface,
curvature can be neglected and the entropy constant derived with a 
plane-parallel calculation (Sect.3.4).  
\subsection{Incident flux}
Consider the point $P$ on the irradiated surface in Fig.1. The incident flux
due to the point source is
\begin{equation}
  \pi F^{inc}_{\mu} = \mu \: \frac{L_{\star}}{4 \pi x^{2}}
\end{equation}
where $\mu = \cos(\chi)$. The resulting incident luminosity is
\begin{equation}
  L^{inc} = 2 \pi R^{2} \int^{1}_{\upsilon_{t}}  \pi F^{inc}_{\mu} d \upsilon 
                   = W L_{\star}
\end{equation}
Here $W = 1/2 \: (1-\sqrt{1-\upsilon_{t}^{2}} \: )$ is the fraction of sky filled
by the target star when viewed from the point source.
Note that
\begin{equation}
  x^{2} = R^{2} + a^{2} - 2 R a \upsilon
\end{equation}
where $\upsilon = \cos \theta$, and 
\begin{equation}
  \mu = \frac{a \upsilon - R}{x}
\end{equation}
where $\mu = \cos (\chi)$.
\subsection{Plane-parallel photosphere}
With the assumption of grey absorption, the temperature distribution 
in the irradiated atmosphere
derived
with Eddington's approximations is (Sobieski 1965) 
\begin{equation}
 B = \frac{\sigma}{\pi} \: T^{4} = 
  \frac{1}{2} \: F_{\mu} \: ( 1 + \frac{3}{2} \tau)
             + \frac{1}{2}\: F^{inc}\: \Psi (\tau;\mu) 
\end{equation}
where
\begin{equation}
  \Psi(\tau;\mu) = 1 + \frac{3}{2} \mu  +
   ( \frac{1}{2 \mu} -  \frac{3\mu}{2}) e^{-\tau/\mu}    
\end{equation}
and $\pi F_{\mu}$ is the (unknown) flux propagating upwards from deep layers
($\tau \rightarrow \infty$). 
\subsection{Inward integrations}
If we define the photosphere to be at $\tau = 2/3$, the photospheric
temperature $T_{ph}(\mu)$ is given by the above formula as
\begin{equation}
 \frac{\sigma}{\pi} \: T_{ph}^{4} = F_{\mu}   
  + \frac{1}{2} F^{inc}_{\mu} \: \Psi(\frac{2}{3};\mu) 
\end{equation}
and the corresponding photospheric pressure is
\begin{equation}
  P_{ph}(\mu)   = \frac{2}{3} \frac{g}{ k_{\rm eff}} 
\end{equation}
where $k_{\rm eff} (T_{ph},g)$, the effective grey absorption coefficient 
defined in Sect.3.2, is assumed to be constant for $\tau < 2/3$.

Because irradiation still affects the temperature stratification for
$\tau > 2/3$, the sub-photospheric radiative gradient $\nabla_{rad}$ is
modified from the familiar diffusion formula. From Eqs.(12) and (13), 
we find that
\begin{equation}
  \frac{d B}{d \tau} = \frac{3}{4} \: F_{\mu} \: \Theta (\tau; \mu) 
\end{equation}
where the correction factor (Rucinski 1969) accounting for irradiation is
\begin{equation}
  \Theta (\tau; \mu) = 1 + \frac{F^{inc}_{\mu}}{F_{\mu}} \: (1 - \frac{1}{3 \mu^{2}})
                                        \:  exp(-\tau/\mu) 
\end{equation}
In order to compute $\Theta$, the additional dependent variable
$\tau$ is required. This is given by  
\begin{equation}
  \frac{d \tau}{d P} = \frac{\kappa}{g}
\end{equation}
with initial 
condition $\tau = 2/3$ at $(P_{ph},T_{ph})$, and where
$\kappa(\rho,T)$ is the Wichita-OPAL opacity (Sect.3.1).

When the inward integration encounters convective instability, standard
mixing length theory with the irradiation-corrected $\nabla_{rad}$ is used
to compute the actual gradient $\nabla$. Note the implicit assumption
that irradiation does not affect $\alpha$, the mixing-length ratio. 

Given $\mu, F_{\mu}, F_{\mu}^{inc}$ and $g$, the initial conditions 
$T = T_{ph}$, $P =  P_{ph}$ and $\tau = 2/3$ allow an inward integration of the 
equations governing the structure of the sub-photosperic layers
as modified by irradiation. But $F_{\mu}$ is not known; it is 
determined by the condition that the entropy constant when convection is 
efficient has the same value as in the corresponding 
integration on the dark side. In effect, $F_{\mu}$ is an eigenvalue determined
by this extra constraint. 

Note that with $F_{\mu} = F^{drk}$ and $F^{inc} = 0$, the above formulae
determine the structure of the photospheric and sub-photospheric layers on
the dark side.    
\subsection{Energy transfer to dark segment}
From the definition of the bolometric albedo $A_{\mu}$ given in Eq.(A.1),
it follows that the outward flux at point P in Fig.1 is 
$F_{\mu}^{+} = F^{drk} + A_{\mu} F^{inc}$ 
Accordingly, the luminosity of the irradiated segment is 
\begin{equation}
  L^{irr} = 2 \pi R^{2} \int^{1}_{\upsilon_{t}}  \pi F^{+}_{\mu} \: d \upsilon  
                    = (1-\zeta) \: L_{\dag} + \langle A_{\mu} \rangle \: L^{inc}
\end{equation}
where $\langle A_{\mu} \rangle$ is the mean bolometric albedo defined by
Eq.(A.2) and $L_{\dag}$ is defined by Eq.(5).
 
Now, if the irradiated segment is in thermal equilibrium, 
\begin{equation}
  L^{irr} = (1-\zeta) \: L_{nuc} + L^{inc} - {\cal L}  
\end{equation}
and if the dark segment is in thermal equilibrium Eq.(6) holds. Combining
these two equilibrium constraints with Eq.(19), we obtain
\begin{equation}
  {\cal L} =  (1-\langle A_{\mu} \rangle) \times \zeta \: L^{inc}  
\end{equation}
as the rate of energy transfer from the irradiated to the dark segment.
For fixed $\langle A_{\mu} \rangle$ and $L_{\star}$, Eqs.(9) and (21) imply
that ${\cal L} \propto \zeta W$, a 
monotonically decreasing function of $a/R$ as expected.

If we use Eq.(21) to eliminate ${\cal L}$ from Eq.(6), we obtain 
\begin{equation}
  L_{nuc} + L^{inc} = L_{\dag} +  \langle A_{\mu} \rangle L^{inc} 
\end{equation}
which is the condition that the irradiated star is in overall thermal 
equilibrium.

Note that the assumption that the target star is in thermal equilibrium
is not necessary in deriving Eq.(21). If the radius $r_{\ell}$ is below
the layers in which the energy transfer occurs and if the layers 
between $r_{\ell}$ and the surface are in thermal equilibrium, then the
above analysis holds with $L_{\ell}$ replacing $L_{nuc}$. Accordingly,
since the layers in question are of negligible mass, the
irradiation modification to stellar structure presented here is also
applicable when $L_{\ell} \ne L_{nuc}$ - i.e., to
rapid phases of stellar evolution.

\subsection{An example}
Consider a detached close binary comprising identical zero-age stars
of mass $0.4 {\cal M}_{\sun}$ separated by $a = 3R_{0}$.
With irradiation neglected, these stars have $L_{0} = 0.215 L_{\sun}$, 
$R_{0} = 0.380 R_{\sun}$, $T_{\rm eff} = 3582K$, and $\log g = 4.880$. 

Treating one of these stars as a point source with $L_{\star} = 
0.215 L_{\sun}$ and neglecting the radius inflation of the target star, we 
calculate ${\cal L}$ as follows:\\

1) Given $T_{\rm eff}$ and $\log g$, the stratification of
the dark-side is obtained with inward integrations as described
in Sect.5.3 - i.e., with $F^{inc} = 0$ and 
$F_{\mu} = F^{drk} = \sigma T_{\rm eff}^{4}/\pi$.
The integration is continued until $T = 10^{5} K$, well into the zone of 
efficient convection (Sect.3.4), so that the adiabat's entropy constant 
$S^{drk}$ is accurately determined.\\

2) Given a value of $\upsilon \in (\upsilon_{t}, 1)$, the incident 
flux $\pi F^{inc}_{\mu}$ is given by
Eq.(8) with $x$ and $\mu$ from Eqs.(10) and (11).       
Then with an initial guess for the eigenvalue $F_{\mu}$, the 
stratification of the irradiated atmosphere is determined as described in
Sect.5.3. The integration is continued until $T = 10^{5} K$ and the
specific entropy constant $S(F_{\mu})$ determined. In general, 
$S(F_{\mu}) \ne S^{drk}$, so that iterative adjustments of $F_{\mu}$ are
required. Iterations with the bisection method determine $F_{\mu}$
with high accuracy.\\

3) Step 2) is repeated for numerous values of $\upsilon$ so that $L^{irr}$
and $\langle A_{\mu} \rangle$ 
given by
Eqs.(19) and (A.2) can be evaluated numerically. Substitution in 
Eq.(21) then gives ${\cal L}$. \\

The result is ${\cal L} = 3.06 \times 10^{-4} L_{\sun}$. Now, with $\zeta = 2/3$
from Eq.(1),
substitution in Eq.(7) gives $\Delta L = 4.59 \times 10^{-4}  L_{\sun}$.   
This is to be compared with the value $ 7.27 \times 10^{-3}  L_{\sun}$
found in Sect.4.2 to be required to inflate the radius of a 
$0.4 {\cal M}_{\sun}$ ZAMS star by $5\%$. Thus, the computed $\Delta L$
falls short by the factor $0.063$.

The energy transferred to the dark segment originates from the energy
incident from the point source. It follows that 
${\cal L} \le L^{inc}$. Now, from Eq.(8), 
$L^{inc} = 6.14 \times 10^{-4} L_{\sun}$, so that ${\cal L}/L^{inc} = 0.498$.
Accordingly, even if this ratio were increased to unity, the resulting
$\Delta L$ would still fall far short of reaching the $5\%$ threshold.
This demonstrates that a non-grey treatment of irradiation cannot
markedly enhance radius inflation.
\section{Consistent models}
In Sect.5.5, the irradiation-induced energy transfer rate ${\cal L}$ is 
computed with the radius of the target star fixed at its value in
the absence of irradiation.
The aim now is to eliminate this inconsistency. This is achieved as 
follows:\\

1) $\Delta L$ is specified and the structure of the equivalent single
star computed as in Sect.4.2. This gives the inflated radius
$R(\Delta L, {\cal M})$ as well as $T_{\rm eff}$ and $g$ of the dark
segment.\\

2) With the specified $a$ and a trial value of $L_{\star}$, the value
of ${\cal L}$ is derived as in Sect.5.5. \\

3) With $\zeta$ from Eq.(1), the value of $\Delta L$ is computed from 
Eq.(7). In general, this differs from the value specified in step 1).
The trial value of $L_{\star}$ in step 2) is therefore iteratively adjusted
to achieve consistency.\\ 

\subsection{A sequence with increasing $L_{\star}$} 

To investigate the dependence of radius inflation on the strength of
the source, a 1-D sequence of consistent models is computed with
fixed $a$ but increasing $L_{\star}$.
Specifically,
the target is a $0.4 {\cal M}_{\sun}$ ZAMS star with the point source
at distance $a = 3 R_{0}$.

The resulting radii $R(L_{\star}; {\cal M},a)$ are plotted in Fig.4. 
As $L_{\star}$ increases, the quantities
$\cal L$ and $\Delta L = {\cal L}/\zeta$ increase, resulting in the plotted 
monotonically increasing radii. For $L_{*} /L_{0} \la 4$, the plot is
accurately linear, giving the approximation
\begin{equation}
  \frac{R}{R_{0}} = 1+ \epsilon \: \frac{L_{*}}{L_{0}}  \;\;\; with \;\;\; 
                                \epsilon = 0.00267  
\end{equation}

Since low-mass stars are also found as companions of luminous primaries,
e.g., sdB + dM binaries,
Fig.4 is extended way beyond $L_{\star} \sim L_{0}$ appropriate for
binaries with M dwarf components. Interestingly, these high $L_{\star}$ 
solutions 
reveal a saturation effect. Thus, beyond $\sim 4\%$ inflation at 
$L_{\star} \sim 60 L_{0}$, a further increase in $L_{\star}$ generates only small 
increases in radius. This arises because, although the energy transfer rate
${\cal L}$ and therefore also $\Delta L$ increase monotonically with 
$L_{\star}$, the ratio ${\cal L}/L^{inc}$ decreases. Thus, as $L_{\star}/L_{0}$ 
increases from
1.0 to 478, the ratio ${\cal L}/L^{inc}$ decreases from 0.49 to 0.016 - i.e., 
the efficiency with which incident energy is converted into energy flow into
the dark segment declines as irradiation increases.
%
%
%
%
\begin{figure}
\vspace{8.2cm}
\includegraphics{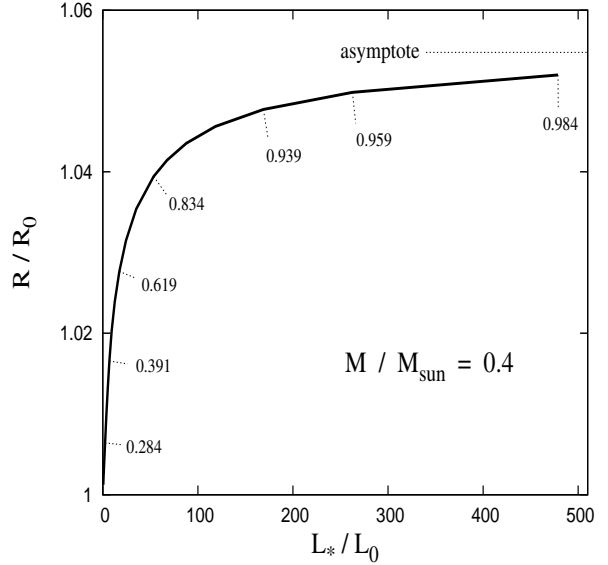}
\caption{Radius inflation of a $0.4 {\cal M}_{\sun}$ ZAMS star due to irradiation
by a point source of luminosity $L_{\star}$ at distance $a = 3 R_{0}$. 
Values of the mean bolometric albedo $\langle A_{\mu} \rangle$ are indicated.
The asymptotic limit corresponds to a radius inflation of $5.48\%$. } 
\end{figure}

This decline in efficiency originates in the structural changes of the
irradiated segment's surface layers as $L^{inc}$ increases. These
result in an increasing mean albedo $\langle A_{\mu} \rangle$ - see Fig.4. 
Thus, the conversion efficiency declines because an increasing fraction of 
$L^{inc}$ is radiated back into space. 

Fig.5 illustrates the structural changes at various points on the irradiated
segment's surface at the onset of the saturation effect. This is taken to be
the point where
$R/R_{0} = 1.04$, corresponding to $L_{\star}/L_{0} = 57.5$ or
$L^{inc}/L_{0} = 1.78$ - i.e., an $O(1)$ perturbation. These $\log P - \log T$
plots show that the irradiated solutions peel away from dark segment solution
at depths that increase with $\mu$. Moreover, the optical depths of
the onset of convection increase from $\tau_{c} = 0.868$ for the dark segment to
$1.03 \times 10^{4}$ when the source is at the zenith $(\mu = 1)$. From
additional such solutions, we find that $\tau_{c} > 5$ when $\mu > 0.236$.
Thus, for $\mu \ga 0.24$, corresponding to $\sim 69\%$ of the
area of the irradiated surface, the incident radiation does not penetrate to
the convection zone, and so the local bolometric albedo $A_{\mu}$ should be
close to unity (Eddington 1926). This partial shielding of the convection zone
by optically-thick radiative zones created by strong irradiation explains why 
$\langle A_{\mu} \rangle    \rightarrow 1$ as 
$L_{\star}/L_{0} \rightarrow \infty$. 

In addition to limiting the amount of radius inflation, the saturation effect
limits the irradiation-induced reduction in nuclear luminosity (Fig.2).
From Fig.4, $R/R_{0} = 1.052$ at $L_{\star}/L_{0} = 479$, and this solution
corresponds to $\Delta L / L_{0} = 0.35$. From Fig.2, we find that the 
resulting drop in $L_{nuc}$ is by $23\%$.

The convergence of solutions to the same adiabat is evident in Fig.5, and
this illustrates the criterion (Sect.5.3) that determines the eigenvalues 
$F_{\mu}$. But the solutions do not overlap when convection is inefficient
or absent. This necessarily implies non-vanishing horizontal pressure
gradients in these layers giving rise to an irradiation-driven circulation 
separate
from that transporting energy between the segments in the deep convective
layers (Sect.4.1). 
This failure of rigorous mechanical equilibrium in the outermost layers
was earlier (Lucy 1967) pointed out for the corresponding gravity-darkening 
problem. Nevertheless, because energy transport by flows in the low density 
surface layers is negligible,
the conclusion that strong irradiation implies 
$\langle A_{\mu} \rangle \rightarrow 1$ is
likely to survive a more rigorous analysis. 
%
%
%
%
%
\begin{figure}
\vspace{8.2cm}
\includegraphics{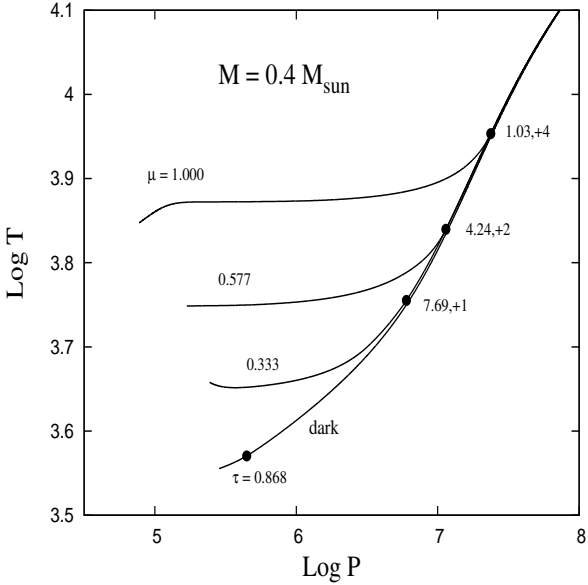}
\caption{Log $P$ - Log $T$ plots for sub-photospheric layers at three points 
on the surface of the irradiated segment as well as for the dark segment. 
The incident direction cosines $\mu$ are indicated.
The filled
circles locate the onset of convection at the indicated optical depths $\tau$. 
These plots correspond to the point $L_{\star}/L_{0} = 57.5, R/R_{0} = 1.04$ in 
Fig.4 where the mean albedo $\langle A_{\mu} \rangle = 0.834$.} 
\end{figure}
\subsection{Asymptotic theory}
Fig.4 strongly suggests that $R/R_{0}$ tends to some {\em finite} 
value as $L_{*}/L_{0} \rightarrow \infty$. 

A clue to the derivation
of this asymptote is found in Fig.5. Prior to the onsets of convection, the
slopes of these plots are the radiative gradients $\nabla_{rad}$. Since
the gradients are near zero - i.e., almost isothermal stratifications,
the radiative fluxes are small, showing that irradiation 
severely inhibits the escape of the internally-generated $L_{nuc}$. 
Extrapolating to
infinite $L_{*}$, we expect zero escape of nuclear energy over the {\em entire}
irradiated surface. Accordingly, in this limit, all energy released by
nuclear burning is radiated by the dark segment, and so the asymptote is
such that
\begin{equation}
  L_{nuc} =  \zeta \: L_{\dag}  
\end{equation}
Substitution in Eq.(6) then gives 
\begin{equation}
  {\cal L} = (1-\zeta) \: L_{nuc}
\end{equation}
confirming that the energy flow rate in the convection zone from the irradiated to
the dark segment is indeed the nuclear burning rate in the irradiated segment.
Finally, substituting in Eq.(7), we find that the required jump in the
luminosity variable in the convection zone of the equivalent single star 
(Sect.4.2) is
\begin{equation}
  \Delta L = \frac{1-\zeta}{\zeta} \: L_{nuc}
\end{equation}
Accordingly, the asymptote is derived by computing a sequence of equivalent 
single stars as in Figs.2 and 3 and iterpolating to find the solution of 
Eq.(26).

For ${\cal M} = 0.4 {\cal M}_{\sun}$, Eq.(26) is satisfied when 
$\Delta L = 0.366 L_{0}$, giving asymptotic inflation of $5.48\%$,
corresponding to $\zeta_{1} = 0.676$.

Throughout this paper, the irradiating source is a point at finite distance 
- Fig.1. In an even simpler model, the source is at infinity 
($a \rightarrow \infty$), in which case the 
target star is irradiated by a parallel beam of intensity $S$, 
and the irradiated fraction $\zeta_{2} = 0.5$. In the limit 
$S \rightarrow \infty$, the 
nuclear burning is again radiated entirely from the dark segment - i.e., 
from exactly 
one hemisphere. The solution of Eq.(26) in this case occurs when  
$\Delta L = 0.633 L_{0}$, giving asymptotic inflation of $10.6\%$.
This increase in asymptotic inflation is a consequence of the increase 
by the factor $(1-\zeta_{1})/(1-\zeta_{2}) = 1.54$ in the surface area from
which escape of $L_{nuc}$ is blocked. 

Note that there is obviously a close
analogy between blocking by the effect of external radiation and the blocking 
that manifests itself as starspots. Thus a single 
${\cal M} = 0.4 {\cal M}_{\sun}$ ZAMS star would experience $10.6\%$ radius 
inflation if $50\%$ of its surface were covered by {\em perfectly black} 
star spots.
\subsection{Equal mass binaries}
Of special interest in Fig.4 is the solution for $L_{\star} = L_{0}$ since this 
corresponds to irradiation by a close companion of equal mass. In this case,
the fractional radius inflation $(R-R_{0})/R_{0}$ is only $0.0026$ 
despite a fractional perturbation  
$L^{inc}/L_{0} = 0.029$. This confirms the earlier conclusion (Sect.4.2) that   
the observed inflations of $\sim 5\%$ cannot be achieved by irradiation from
similar mass companions.

To check the generality of this conclusion for low-mass binaries with
main sequence components, solutions for ${\cal M}/ {\cal M}_{\sun}
= 0.2 (0.1) 0.6$ are reported in Table 1. In each case,
$L_{\star} = L_{0}({\cal M})$ and $a = 3 R_{0}$. These further support the 
conclusion 
that inflation-induced inflation of the components' radii are too small 
by more than an order of magnitude to explain the values inferred from
observational data (Sect.1).

\section{Summary}
The aim of this paper is to quantify the radius inflation of the components
of low-mass binaries due to their mutual irradiation. The motivation is to
test whether this effect can significantly contribute to explaining 
the well-established disagreements between the radii derived from
eclipsing binaries and the radii of stellar models (Sect.1).   

In order to predict the inflationary effect of irradiation, the classical
stellar structure boundary condition  
of no inwardly-directed radiation at $\tau = 0$ must be abandoned.
In its place, a boundary condition is derived by considering irradiation
by a point source (Sect.2, Fig.1) and exploiting a 
treatment of the temperature stratification in the surface layers 
previously-developed for photometric analyses of eclipsing binaries (Sect.5).
  
A crucial element in the structure of the irradiated star is the energy transfer
that occurs within the convection zone from the irradiated to the dark segment 
(Sect.5.4). This transfer is induced by the irradiation, and its rate derives
from the constraint of zero entropy gradient within the simply-connected zone 
of efficient
convection which lies beneath the irradiated and non-irradiated surface 
layers. The luminosity increment within the dark segment resulting from 
this energy
transfer inflates the star's radius (Sect.4). A consistent model is 
achieved when
the energy intercepted by the inflated star induces exactly the energy 
transfer required to produce the inflated radius (Sect.6).

A sequence of consistent models appropriate for low-mass
zero-age close binaries is reported in Table 1. The inflated radii are 
$\la 0.4\%$, considerably
smaller than the claimed discrepancies (Sect.1), and so unlikely to be
eliminated by introducing such refinements as a finite-sized irradiation 
source or a non-grey treatment of the reflection effect.  
Accordingly, the observed inflated radii remain unexplained.

\begin{table}

\caption{Irradiated low-mass ZAMS stars. }

\label{table:1}

\centering

\begin{tabular}{c c c c c}

\hline\hline

  $ {\cal M}/{\cal M}_{\sun} $    &  $ \log L_{\star}/L_{\sun}$  &  $a/R_{\sun}$ 
&  $\langle A_{\mu} \rangle$  & $\Delta R / R_{0}  (\%)$    \\

\hline
\hline

     0.2 &  -2.215   & 0.699   & 0.30  & 0.20    \\

     0.3 &  -1.894   & 0.930   & 0.27  & 0.21    \\

     0.4 &  -1.668   & 1.141   & 0.26  & 0.26    \\

     0.5 &  -1.424   & 1.395   & 0.28  & 0.37    \\

     0.6 &  -1.148   & 1.684   & 0.32  & 0.39    \\

\hline
\hline

\end{tabular}

\end{table}
\acknowledgements

I thank the referee, D.Homeier, whose pertinent comments improved the
paper.
\appendix  

\section{Bolometric albedo}
In deriving photometric elements of eclipsing binaries with
a cool component, the reflection effect is commonly taken into account 
with a bolometric albedo of 0.4-0.5 following Rucinski (1969). 
As emphasized by Rucinski, the reflection effect in a photometric analysis
relates to the enhancement of the flux $F^{+}_{\mu}$ from the irradiated surface
above that on the dark side $F^{drk}$, since an isotropic change 
due to the changed
internal structure (Fig.2) is not observed as a photometric perturbation.
Accordingly, at point $P$ in Fig.1, the bolometric albedo is 
defined to be 
\begin{equation}
  A_{\mu} = \frac{F^{+}_{\mu} - F^{drk}}{F^{inc}_{\mu}} 
\end{equation}
and, since $F_{\mu}^{+} = F_{\mu} + F^{inc}$, its determination is trivial once 
the eigenvalue $F_{\mu}$ has been found (Sect. 5.5).

Because the $\mu$-dependence of $A_{\mu}$ is not readily inferred in a
photometric analysis, it is
useful to define a mean albedo. A natural choice is the mean obtained by
weighting $A_{\mu}$ by $F^{inc}_{\mu}$ and averaging over the irradiated surface.
Thus, we define  
\begin{equation}
  \langle A_{\mu} \rangle  = 
   \int_{\upsilon_{t}}^{1}  A_{\mu} F^{inc}_{\mu} d \upsilon \:/
                          \int_{\upsilon_{t}}^{1} F^{inc}_{\mu} d \upsilon  
\end{equation}
Substituting $A_{\mu}$ from Eq.(A.1) and making use of Eqs.(1),(5),(9) and (19),
we find that
\begin{equation}
  \langle A_{\mu} \rangle = (L^{irr} - L^{irr}_{0}) \:/\: L^{inc} 
\end{equation}
where $L^{irr}_{0} = (1-\zeta) \: L_{\dag}$ is the luminosity of
the irradiated segment when $F^{+}_{\mu} = F^{drk}$.  

In the limit $a/R \rightarrow \infty$, we have $\upsilon \rightarrow \mu$
so that $\langle A_{\mu} \rangle \rightarrow  \bar {A}_{\mu}$, where 
\begin{equation}
  \bar {A}_{\mu}   = 
   \int_{0}^{1}  A_{\mu} F^{inc}_{\mu} d \mu \:/
                          \int_{0}^{1} F^{inc}_{\mu} d \mu  
\end{equation}
If we further suppose that $ L^{inc} \ll L_{0}$, then  $\bar {A}_{\mu}$
is just a function of the stellar atmosphere parameters
$T_{\rm eff}$ and $g$.

\end{document}